\documentclass{PoS}

\title{Recent Developments in CMS Fast Simulation}

\ShortTitle{Recent Developments in CMS Fast Simulation}

\author{\speaker{Sezen Sekmen}\\
        (On behalf of the CMS Collaboration) \\
        Kyungpook National University, Daegu, South Korea \\
        E-mail: \email{ssekmen@cern.ch}}

\abstract{CMS has developed a fast detector simulation package, which serves as a fast and reliable alternative to the detailed GEANT4-based (full) simulation, and enables efficient simulation of large numbers of standard model and new physics events.  Fast simulation becomes particularly important with the current increase in the LHC luminosity.  Here, I will discuss the basic principles behind the CMS fast simulation framework, and how they are implemented in the different detector components in order to simulate and reconstruct sufficiently accurate physics objects for analysis.  I will focus on recent developments in tracking and geometry interface, which improve the flexibility and emulation performance of the framework, and allow a better synchronization with the full simulation.  I will then show how these developments have led to an improved agreement of basic analysis objects and event variables between fast and full simulation.}

\FullConference{38th International Conference on High Energy Physics\\
		3-10 August 2016\\
		Chicago, USA}

\begin{document}

\section{Introduction}

The vast spectrum of CMS physics analyses rely on large numbers of simulated events in order to obtain the most definitive physics results.  CMS has developed a rigorous GEANT4~\cite{Agostinelli:2002hh}-based full detector simulation framework (FullSim) that simulates the traversing particle interactions within the CMS detector in great detail.  FullSim is dominantly used for standard model processes, which need to be accurately modelled as signals in measurement studies or backgrounds in new physics searches.  However FullSim accuracy costs minutes of simulation time per event.  To address the issue, CMS has also developed a mainly parametric fast simulation (FastSim) framework~\cite{Abdullin:2011zz}, which reduces the simulation time by a factor of $\sim$100 and simulation$+$reconstruction time by a factor of $\sim$20, yet provides a reliable alternative that reproduces FullSim with $\sim$10\% accuracy.  

FastSim is an integral part of the CMS software framework and physics studies.  It is currently used for simulating large supersymmetry signal scans, some exotic signal scans, samples with varied systematics for top quark measurements, and private samples for MSc and PhD theses.  As increasing LHC luminosity and pile up will require ever higher numbers of events, FastSim is soon expected to find wider usage, starting with the upgrade studies.  Therefore, a team of $\sim$15 developers constantly work to maintain and improve FastSim in CMS.  In the following, I will give a brief overview of FastSim and the principles that make it fast; summarize the recent developments in FastSim tracking and calorimetry; and present latest results from FastSim performance in comparison to FullSim.

\section{Brief overview of FastSim}

Detector simulation can be considered in three main steps: 1) Simulation, where interactions of a traversing particle with the detector and the resulting energy depositions in the detector are simulated; 2) Digitization, where energy depositions in the detector are converted into digital signals; and 3) Reconstruction, where the digital signals are subsequently reconstructed into objects for physics analysis.  Detector hits after the simulation and reconstruction steps are called SimHits and RecHits respectively.  In CMS, main difference between FullSim and FastSim arises in the simulation step.  FullSim uses the exact detector geometry, tracks particles in small steps, and uses detailed models for material interactions, whereas FastSim uses a simplified geometry with infinitely thin material layers and simple analytical material interaction models that are parametrized and tuned to agree with FullSim.  For the digitization step, both FullSim and FastSim do a detailed emulation of detector electronics and trigger, with small exceptions in FastSim.  In the reconstruction step, FullSim employs the standard event reconstruction used for reconstructing the real CMS data. FastSim uses standard reconstruction for calorimetry and muon systems, but a simplified reconstruction for tracking, based on smearing and truth information, in order to reduce CPU time.

\section{FastSim tracking and recent developments}

FastSim tracking starts with the simulation of SimHits in the tracker.  For simplicity, CMS tracker geometry is approximated by infinitely thin cylinders and disks.  Materials reside on the surfaces of geometry layers, and material simulation is done by assigning thickness in radiation lengths to the cylinders.  Particles are propagated within a simplified B field, and particle-tracker material interactions occur in each layer.  Next, particle intersections in the simplified geometry are projected onto the realistic tracker geometry, and the entry and exit points of a particle trajectory are calculated, which gives the SimHits.  

Emulation of tracker RecHits is a simplified process in FastSim based mainly on smearing of the SimHits.  For the pixel detector, which occupies the innermost layers of the CMS tracker, a detailed position smearing is performed based on position resolution templates obtained from a dedicated simulation by PIXELAV~\cite{Swartz:2003ch}, dependent on local coordinates, pixel type, incident angle, and number of pixels that are hit.  Hits can merge when their charge deposits overlap.  In FullSim, tracking allows to share merged hits between tracks.  The probability of two hits merging in FastSim given local $\eta$ and distance between hits was recently simulated also using PIXELAV.
For the strip detector surrounding the pixels, hit positions are determined with a simple Gaussian smearing.  Recently, a new FastSim tracking RecHit producer package was developed to emulate tracking RecHits in a modular and systematic fashion.  This modularity will allow FastSim to easily adapt to any (upgrade) geometry.  The package consists of plugins configured per detector ID by python configuration, and allows to activate different levels of detail when relevant.  

\begin{figure}[htbp]
\begin{center}
\includegraphics[width=0.9\textwidth]{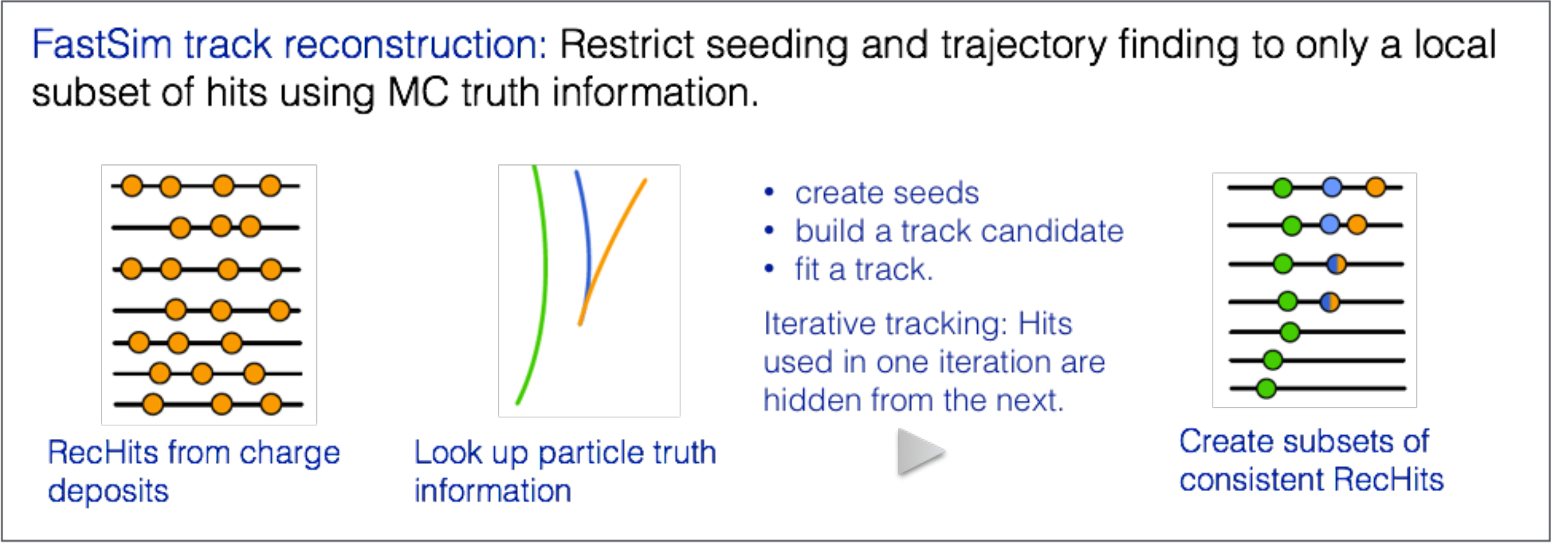}
\caption{FastSim track finding principle.}
\label{fig:fasttrack1}
\end{center}
\end{figure}

\begin{figure}[htbp]
\begin{center}
\includegraphics[width=0.9\textwidth]{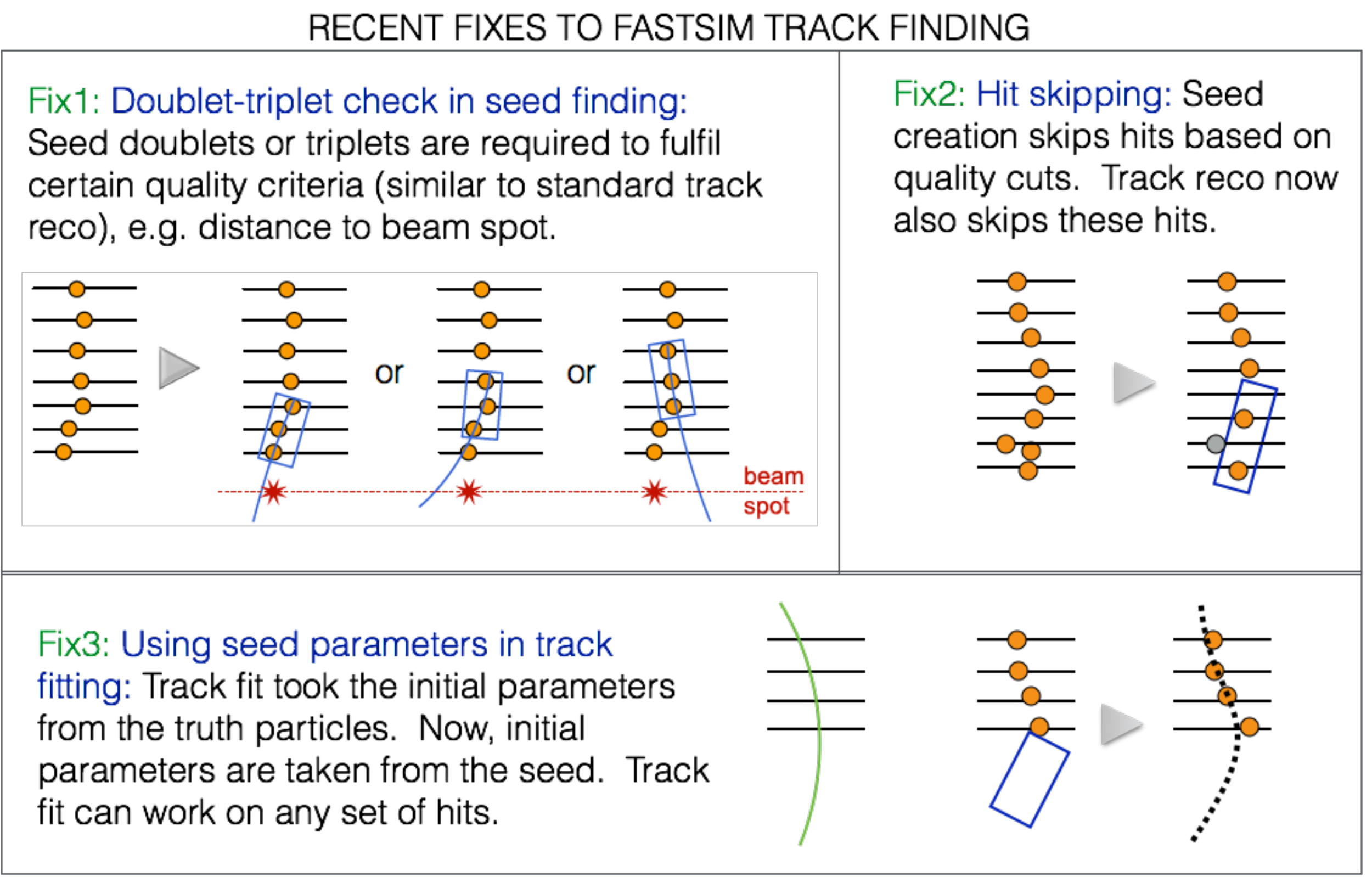}
\caption{Summary of recent modifications to FastSim track finding.}
\label{fig:fasttrack2}
\end{center}
\end{figure}

\begin{figure}[htbp]
\begin{center}
\includegraphics[width=0.35\textwidth]{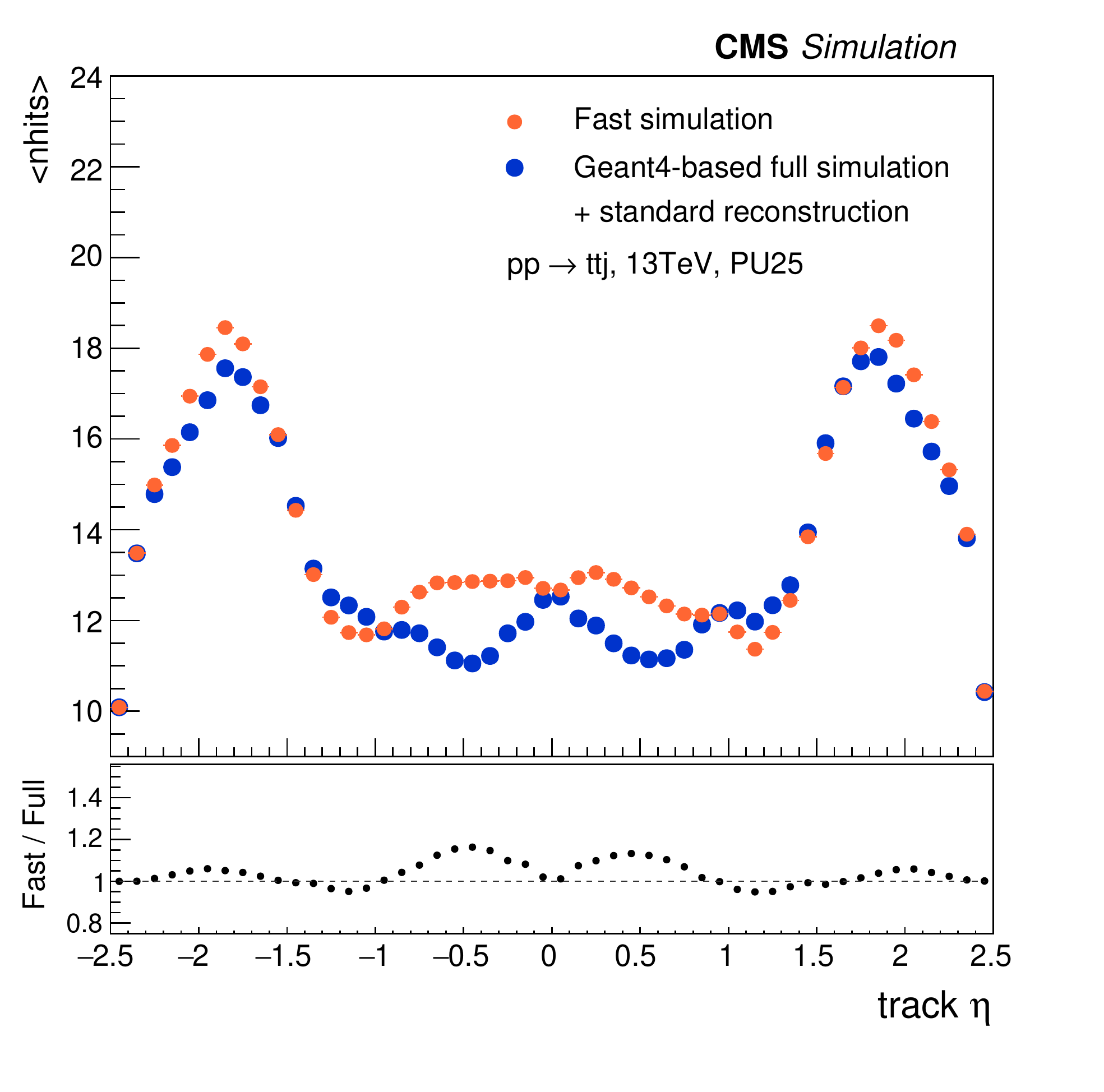} 
\includegraphics[width=0.35\textwidth]{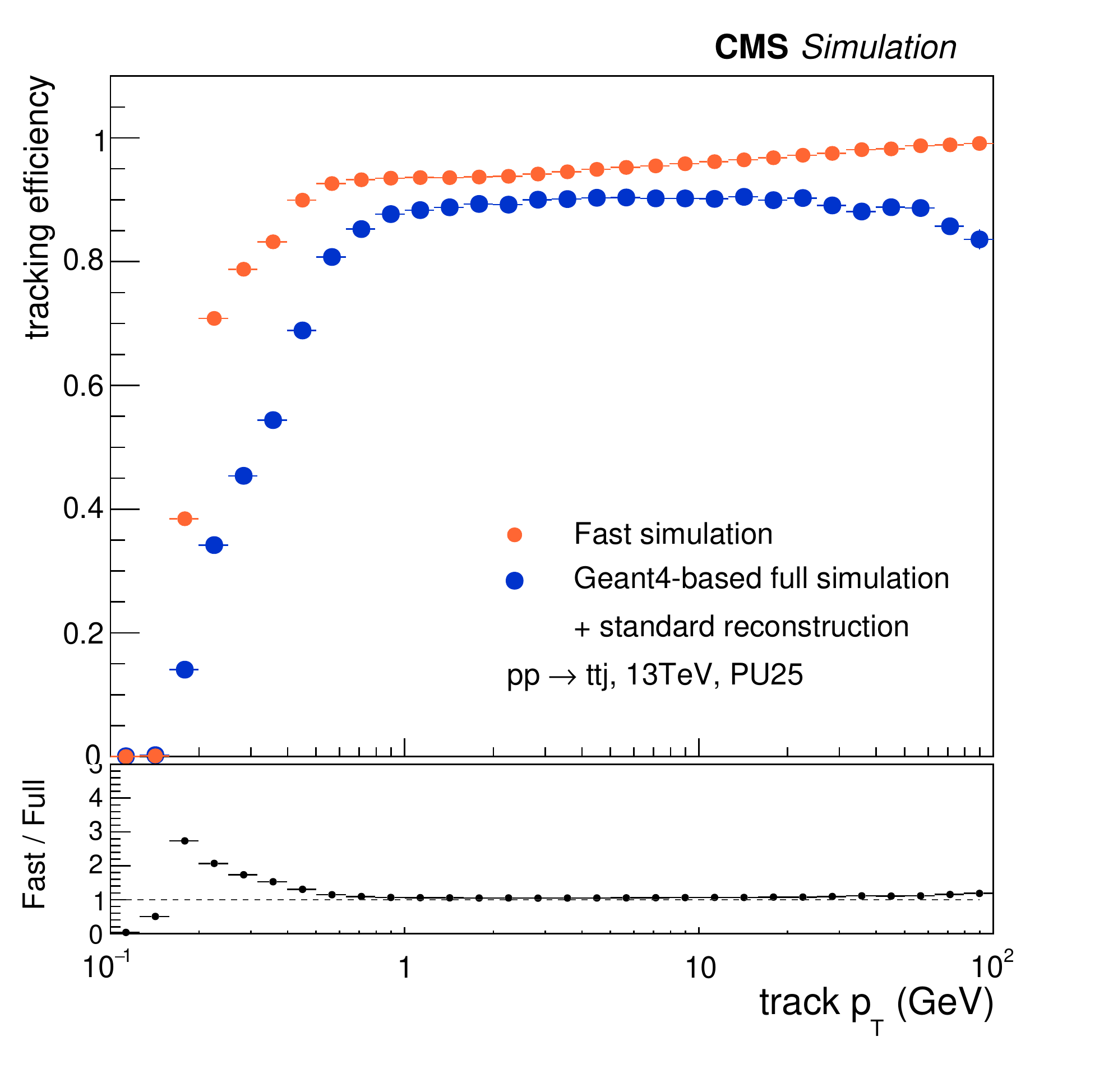} 
\caption{average number of RecHits as a function of track $\eta$ (left) and track finding efficiency as a function of $p_T$ (right) between FastSim and FullSim after the modifications.}
\label{fig:trackcomp}
\end{center}
\end{figure}

Next, tracking RecHits are used for reconstructing tracks.  FullSim considers combinations of hits from a nearly infinite number of hit permutations created by charged particle trajectories, bent by the B field.  FastSim restricts track seed and trajectory finding to only a local subset of hits using MC truth information, as shown in Figure~\ref{fig:fasttrack1}.  The skipping of hit permutations yields one of the major speedup compared to standard reconstruction.  The whole tracking sequence (seeding $\rightarrow$ track finding/trajectory building $\rightarrow$ track fitting) is performed iteratively, where RecHits used in one iteration are not used in the next ones.  Recently, FastSim track finding algorithm was modified to become more compatible with the full track finding.  Figure~\ref{fig:fasttrack2} summarizes these modifications.  Figure~\ref{fig:trackcomp} compares average number of RecHits as a function of track $\eta$ (left) and track finding efficiency as a function of $p_T$ (right) between FastSim and FullSim after the modifications.  These fixes have significantly improved FastSim agreement with FullSim.

\section{FastSim calorimetry and recent developments}

Particles passing through the CMS calorimeter system generate electromagnetic (EM) and hadronic showers.  In FastSim, these showers are simulated by first finding energy dependent parametrizations of the shower properties such as the shower starting point and shower shapes, then fitting these parametrizations to FullSim to extract the best parameters.  Then, energy spots are sampled within the showers and converted into SimHits.  Next, standard reconstruction is applied to obtain calorimeter RecHits.  

\begin{figure}[htbp]
\begin{center}
\includegraphics[width=0.24\textwidth]{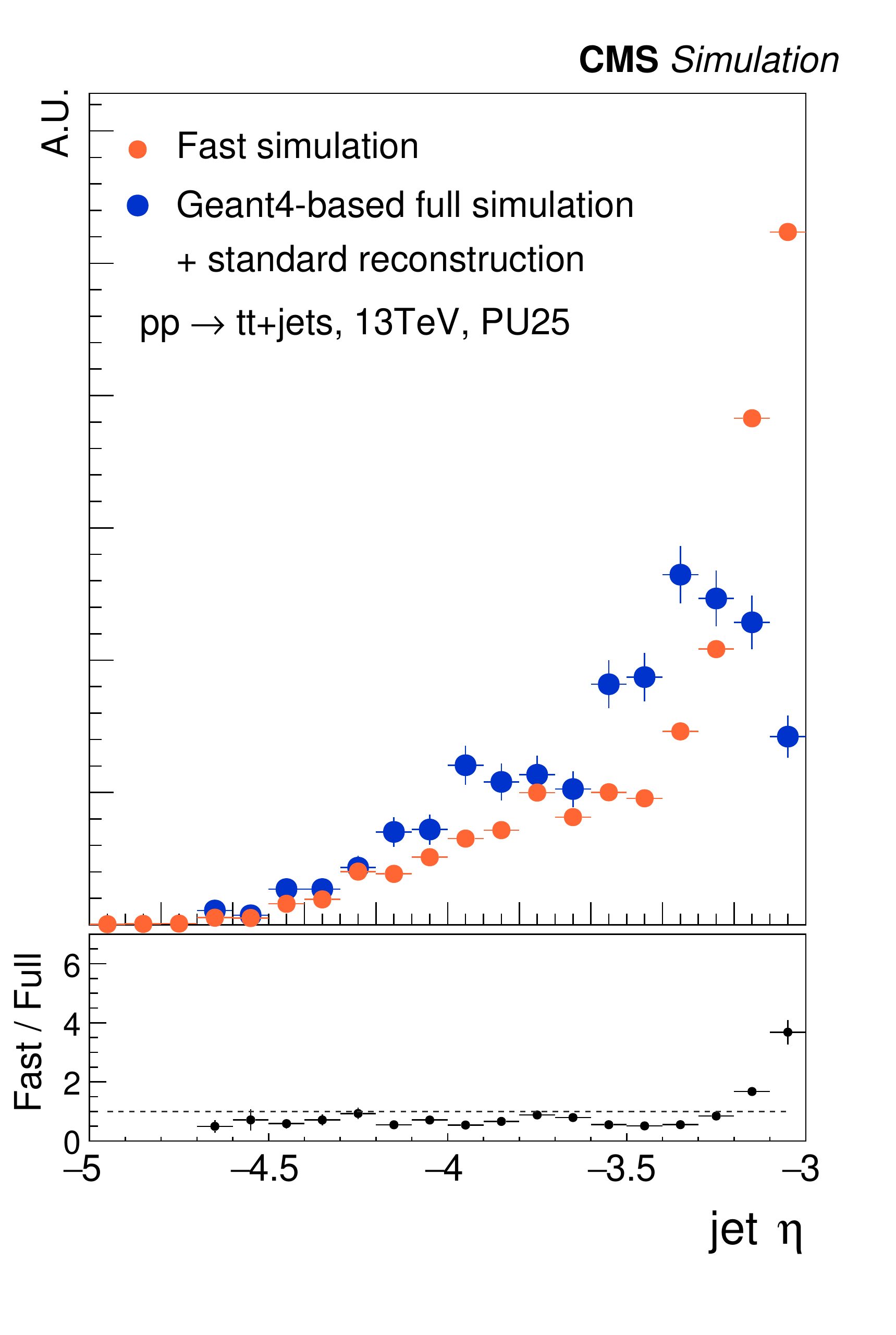}
\includegraphics[width=0.24\textwidth]{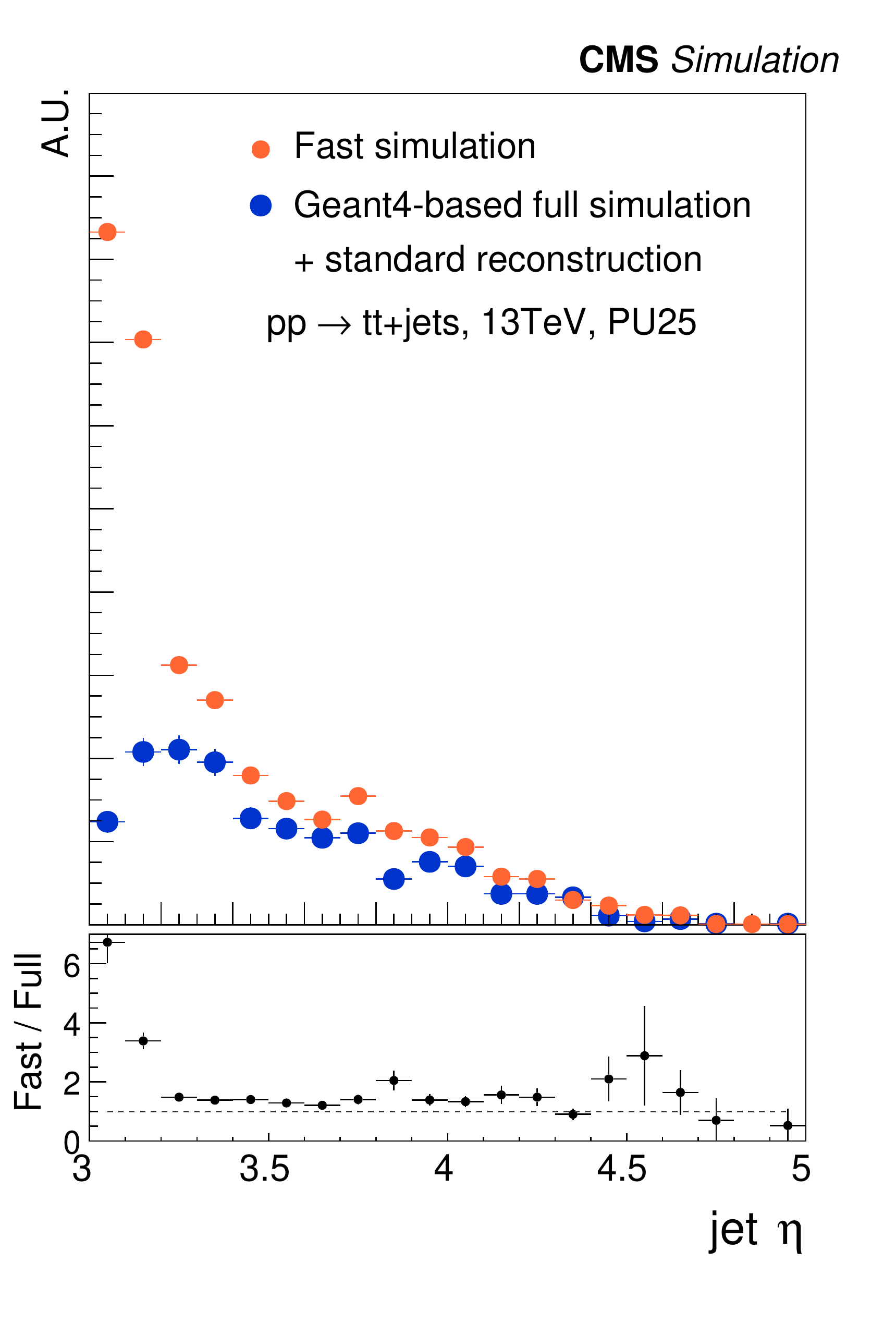} 
\includegraphics[width=0.24\textwidth]{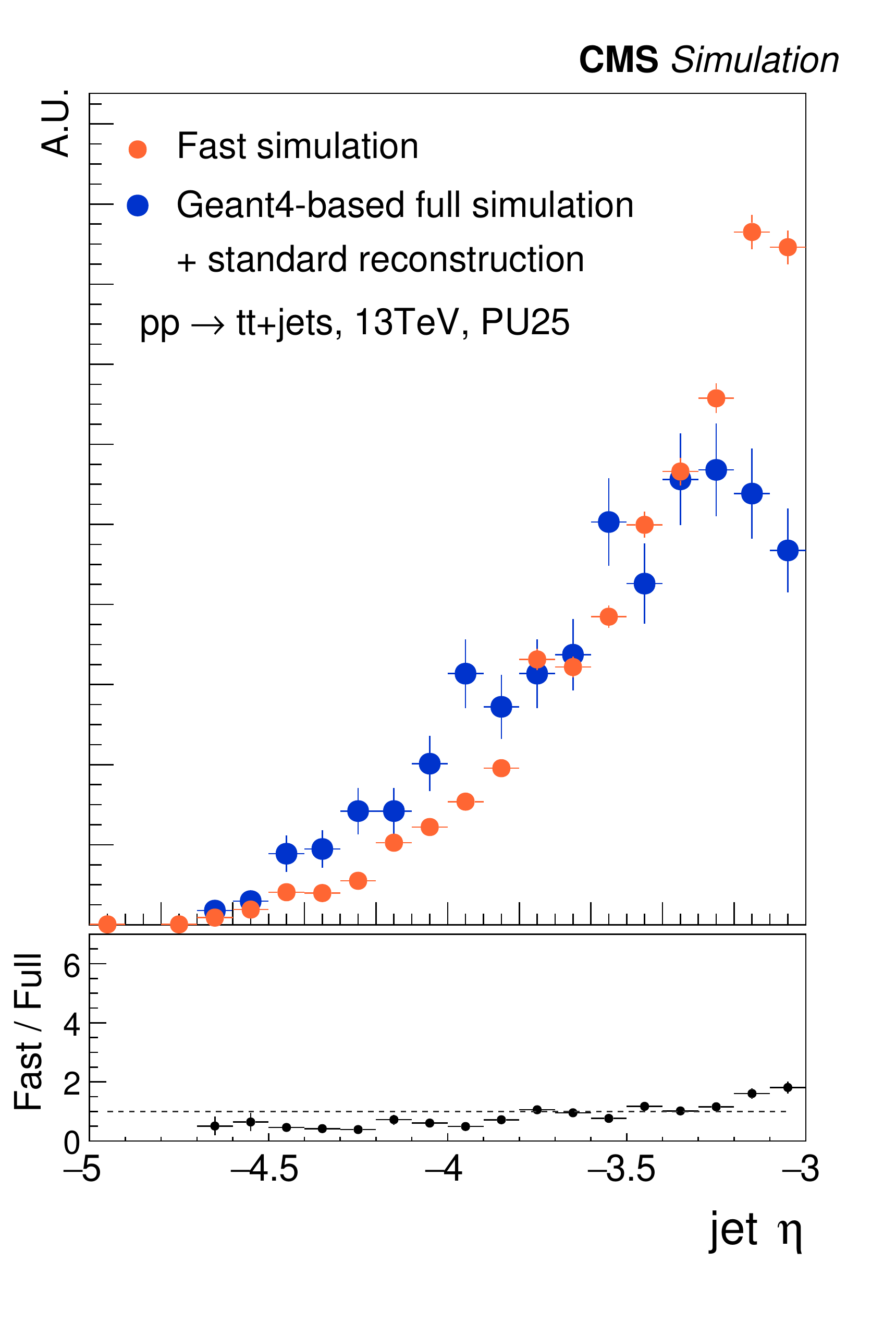} 
\includegraphics[width=0.24\textwidth]{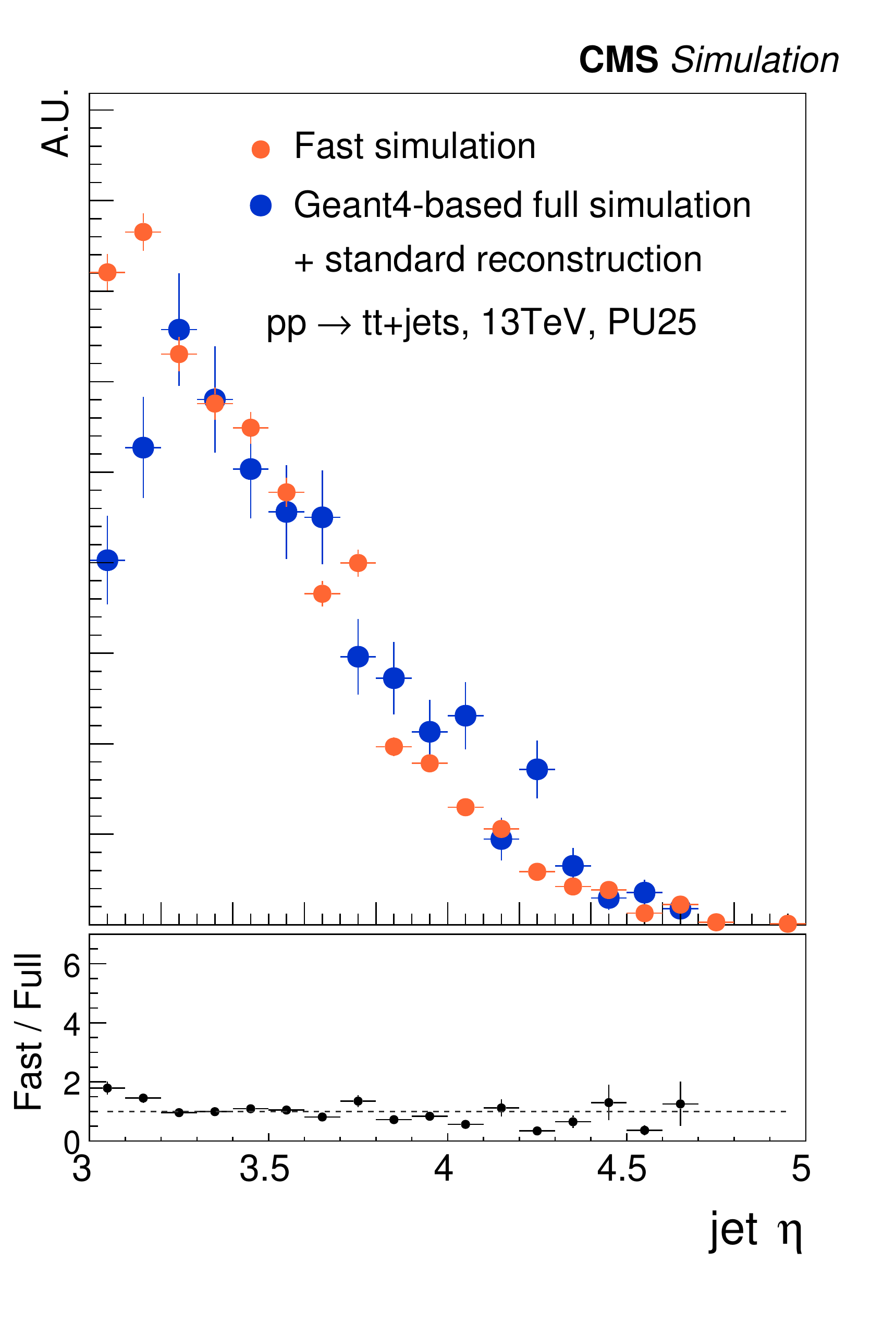} 
\caption{FastSim-FullSim comparison of forward jet $\eta$ distributions before (left) and after (right) recent improvements in the hadron forward calorimetry shower library.}
\label{fig:HFchange}
\end{center}
\end{figure}

However, this method does not yield sufficiently accurate results for the forward calorimetry region.  Therefore, showers in the forward region are directly taken from a shower library previously simulated using GEANT4.  Here, showers are classifed according to particle type, energy and $\eta$.  Recently, a further improvement was made by applying FastSim-specific correction factors to take into account the incomplete modelling of material in front of the hadron forward calorimeter.  Figure~\ref{fig:HFchange} shows the forward jet $\eta$ distributions before and after the corrections, where a significant improvement can be observed.

\section{FastSim objects performance and validation}

\begin{figure}[htbp]
\begin{center}
\includegraphics[width=0.32\textwidth]{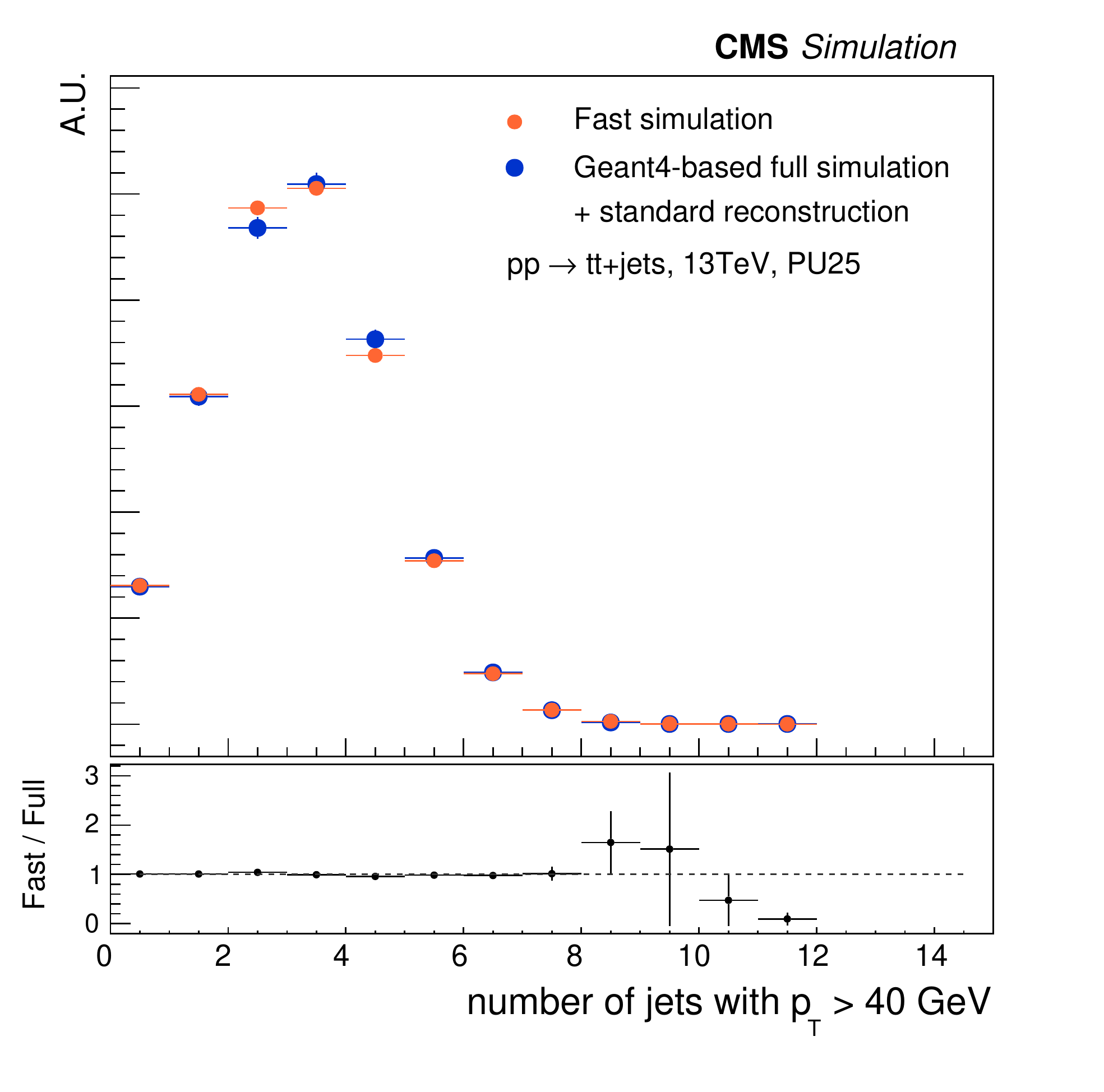}
\includegraphics[width=0.32\textwidth]{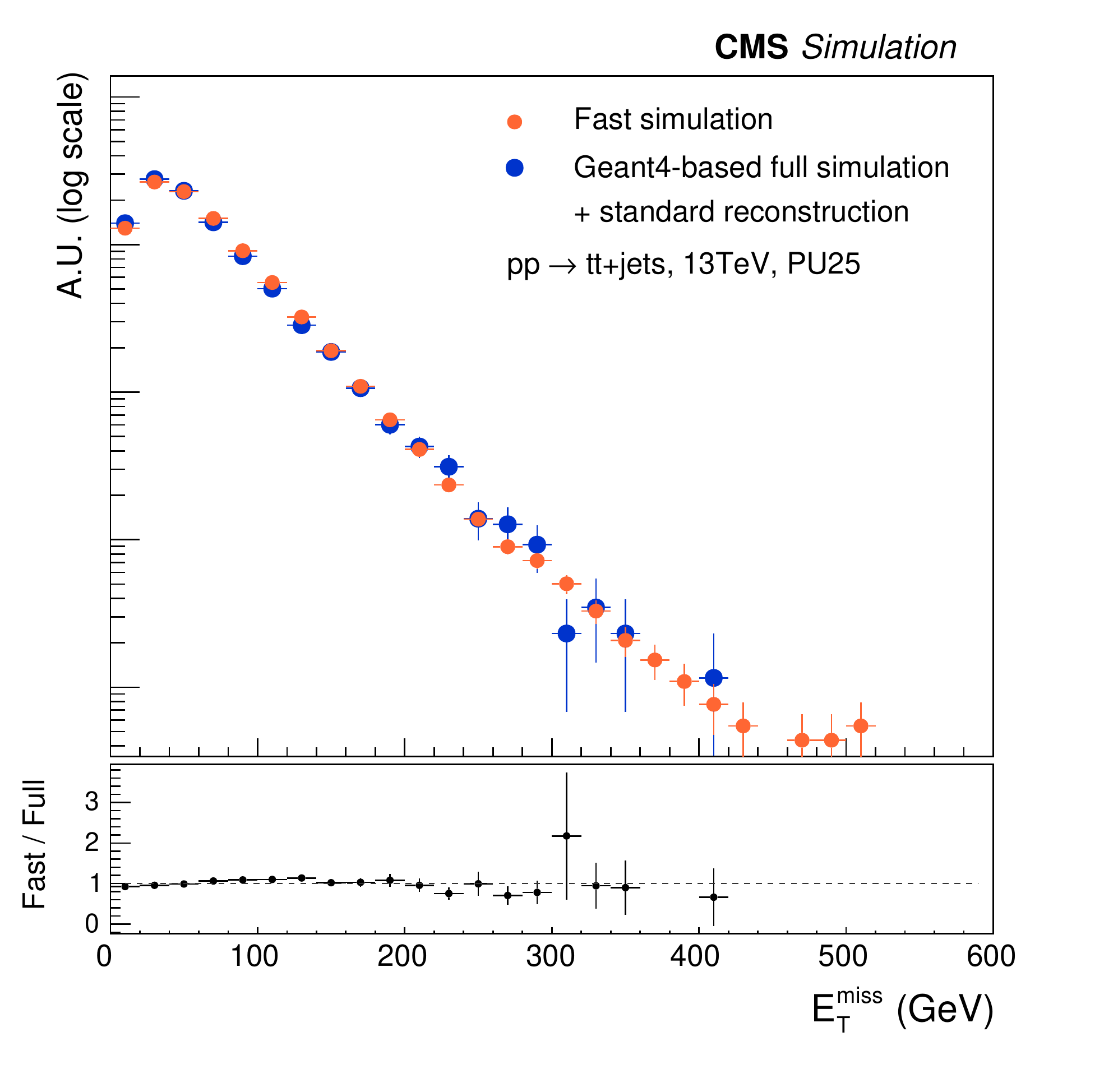} 
\includegraphics[width=0.32\textwidth]{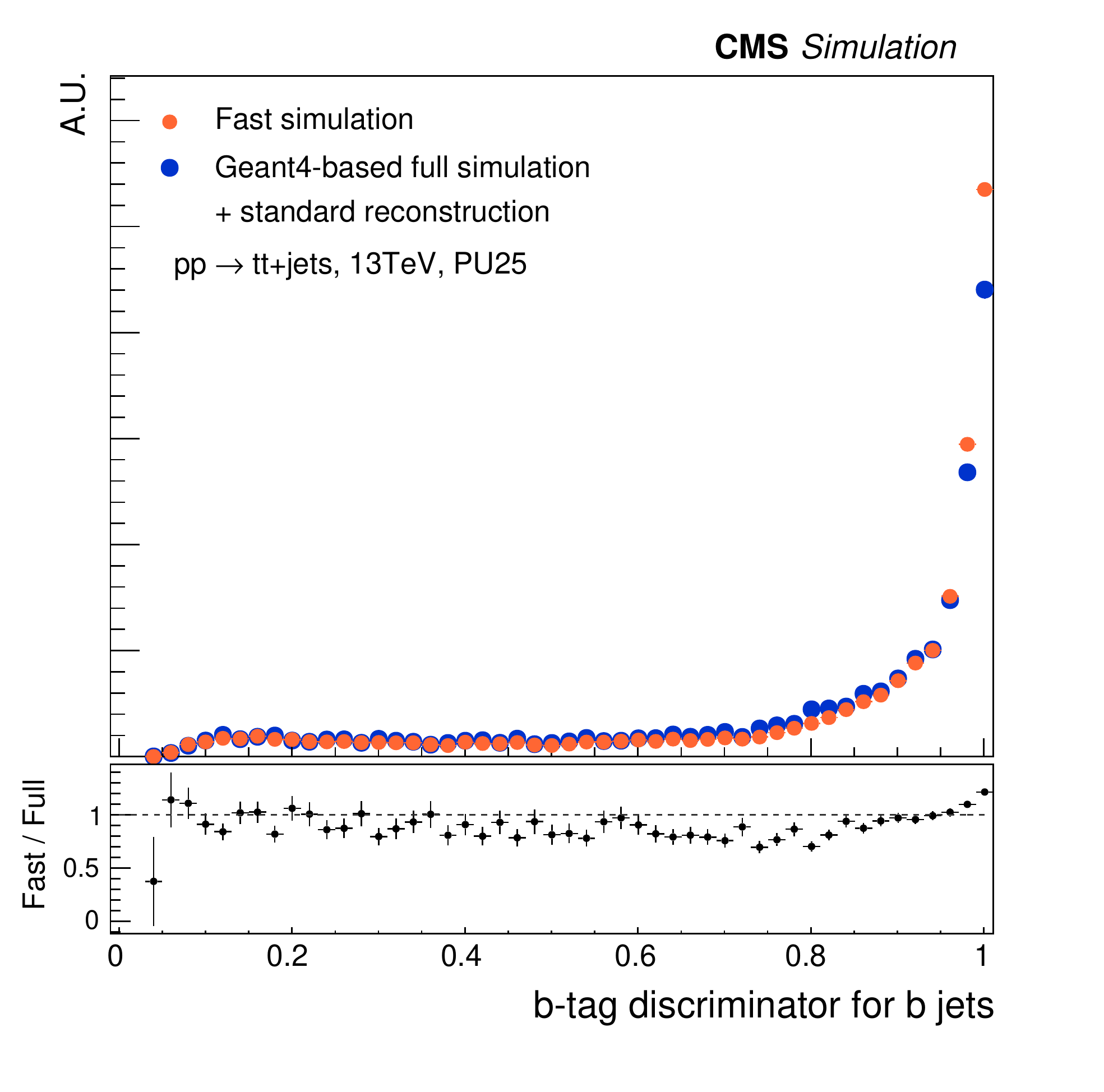} \\ 
\includegraphics[width=0.32\textwidth]{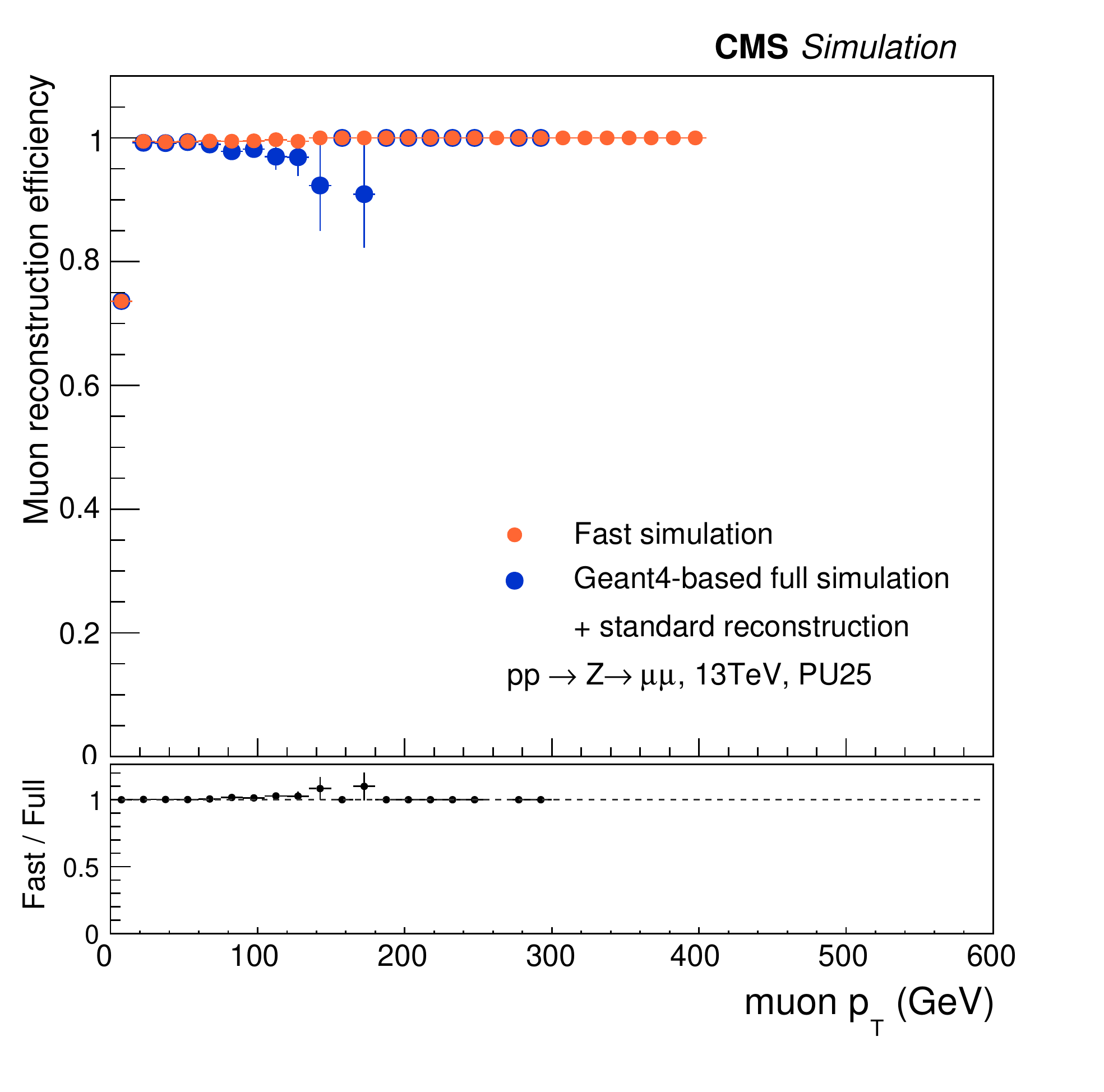} 
\includegraphics[width=0.32\textwidth]{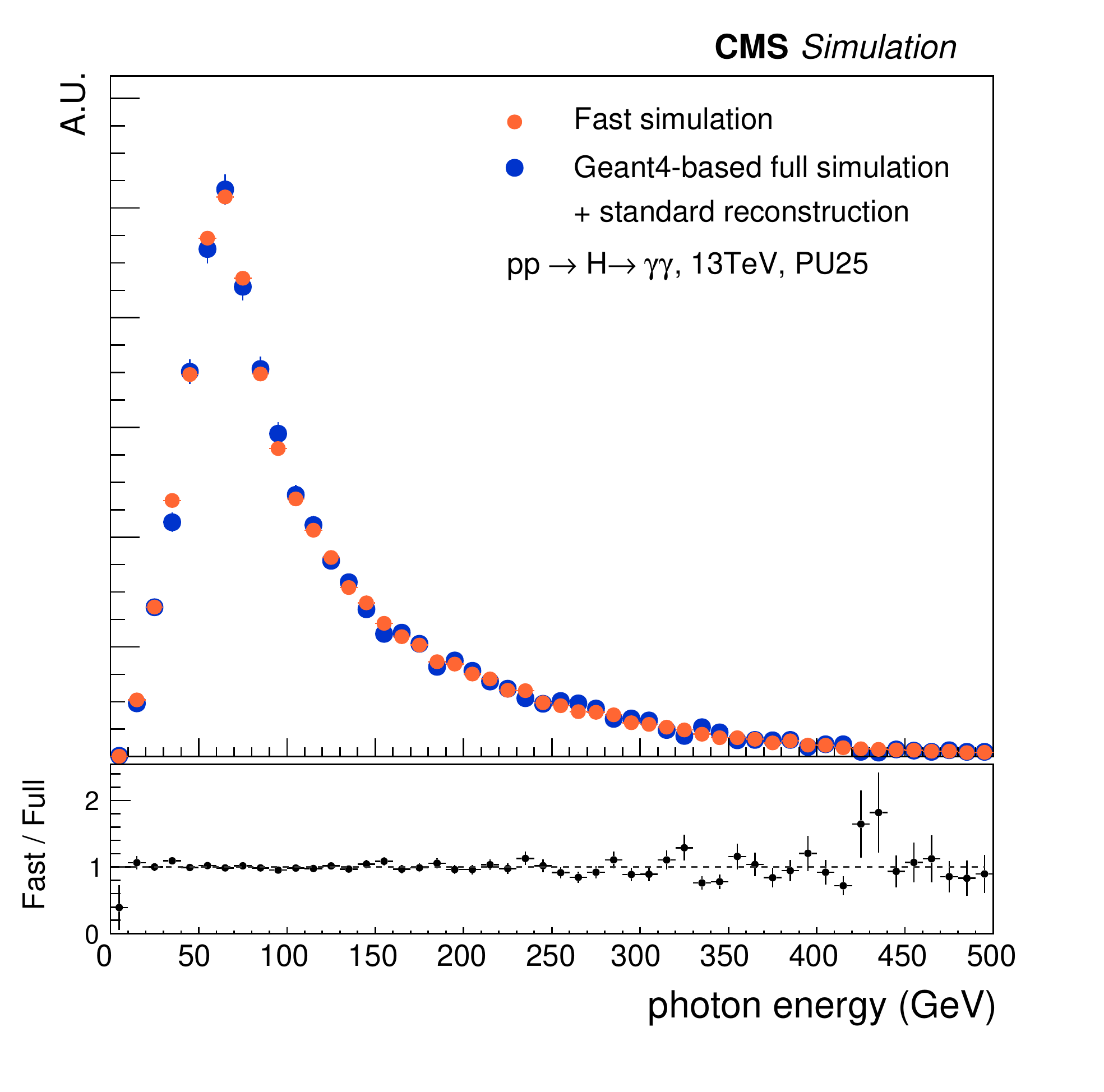} 
\includegraphics[width=0.32\textwidth]{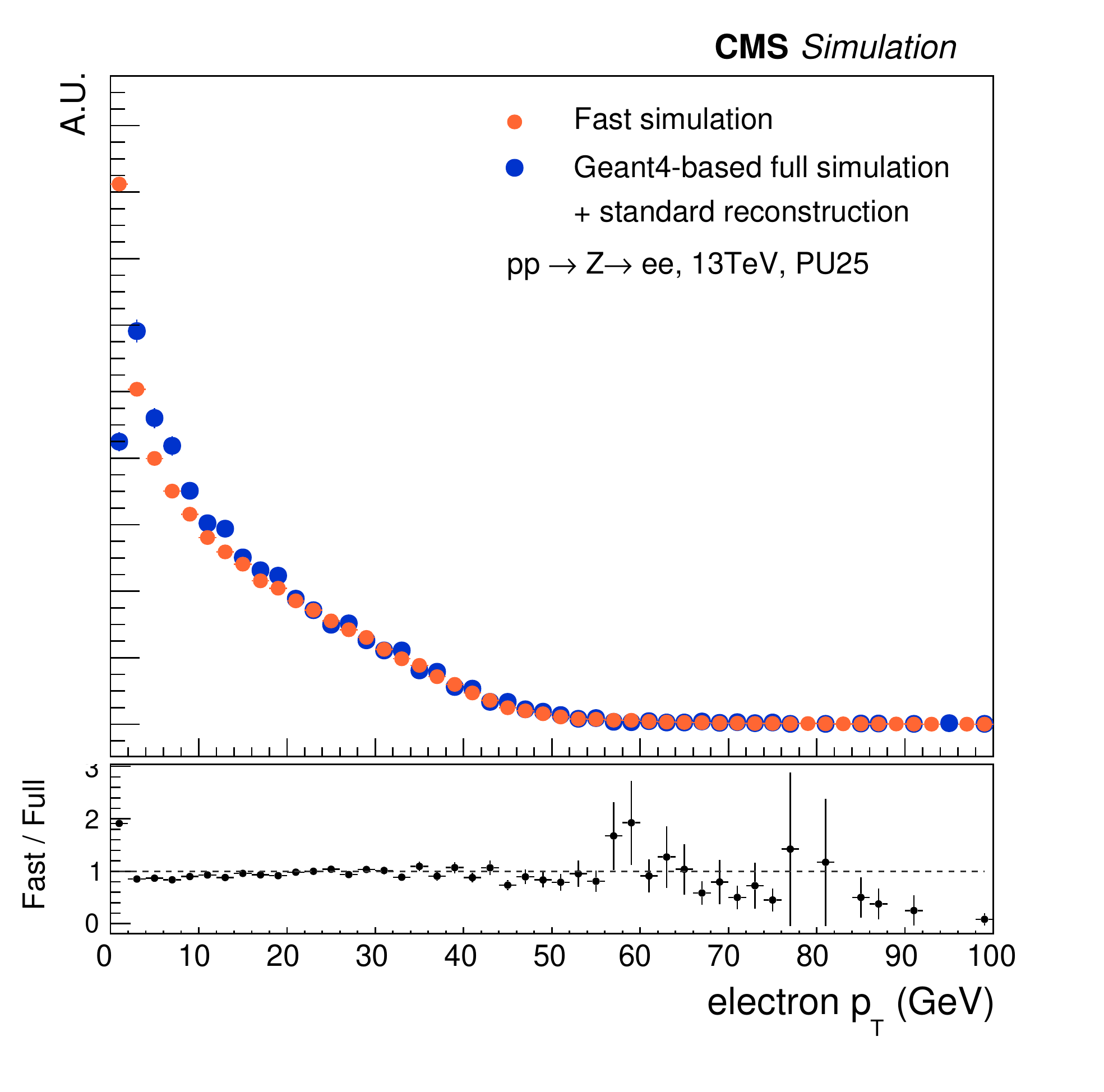} 
\caption{FastSim-FullSim comparison for selected high level object distributions.  See text for details.}
\label{fig:validation}
\end{center}
\end{figure}

Performance of FastSim is validated regularly within the official CMS software release validation framework.  Distributions for low level variables and higher level objects are compared to FullSim.  Figure~\ref{fig:validation} shows some example FastSim-FullSim comparison for selected high level objects.  Overall, distributions in FastSim agree with FullSim within $\sim 10\%$.

Top left plots shows number of jets.  Jets are reconstructed with the particle flow algorithm, using the tracker in $|\eta| < 2.5$ but rely exclusively on calorimeters outside.  Difference with respect to FullSim would arise from tracking and calorimetric shower simulation.  Basic jet properties, jet energy scale, resolution, etc. perform excellent in FastSim.  Jet composition performance, which is poorer, will be improved by the current developments in shower modelling,  Top center plot shows b jet tag discriminator value for truth b jets.  FastSim uses the same b jet tagging algorithms with FullSim.  Despite the differences in FastSim tracking and vertexing, overall performance is very good, and was improved by the latest tracking developments.  
Top right plot shows missing transverse energy ($E_T^{miss}$), whose well-performance is especially important, as all supersymmetry signals are generated using FastSim.  $E_T^{miss}$ is also the "golden observable" where any significant mis-modelling of any object and the whole event will show up.  Overall, a very good agreement is observed.  $E_T^{miss}$ resolution will be further improved by ongoing calorimetry shower developments.  Next, in the bottom left plot, we see the good performance of muon efficiency versus muon $p_T$.  Main difference in muons comes from tracking.  Finally, bottom center and right plots show photon energy and electron $p_T$.  The main difference from FullSim comes from EM showers for photons, and EM showers plus tracks for electrons.  Energy scale, resolution, identification and isolation variables are well-reproduced, but cluster shapes, and track-cluster matching for electrons can be further improved.

Overall, FastSim has established itself as a fast and alternative method of detector simulation in CMS.  Work continues to improve its performance both on the tracking and calorimetry fronts.  Meanwhile, developments are underway to ensure FastSim readiness for the upcoming Phase1 and Phase2 detector upgrades.

\section*{Acknowledgements}

I would like to thank all developers in the CMS FastSim team for their constant, dedicated efforts, and CMS offline and computing group for their support.


\begin{thebibliography}{99}

\bibitem{Agostinelli:2002hh}
  S.~Agostinelli {\it et al.} [GEANT4 Collaboration],
  ``GEANT4: A Simulation toolkit,''
  Nucl.\ Instrum.\ Meth.\ A {\bf 506} (2003) 250.
  doi:10.1016/S0168-9002(03)01368-8

\bibitem{Abdullin:2011zz}
  S.~Abdullin {\it et al.} [CMS Collaboration],
  ``The fast simulation of the CMS detector at LHC,''
  J.\ Phys.\ Conf.\ Ser.\  {\bf 331} (2011) 032049.
  doi:10.1088/1742-6596/331/3/032049

\bibitem{Swartz:2003ch}
  M.~Swartz,
  ``CMS pixel simulations,''
  Nucl.\ Instrum.\ Meth.\ A {\bf 511} (2003) 88.
  doi:10.1016/S0168-9002(03)01757-1


\end{thebibliography}
\end{document}